\documentclass[twoside]{photon2007}
\usepackage[latin1]{inputenc}
\usepackage[dvips]{graphicx,epsfig,color}
\usepackage{wrapfig,rotating,psfrag}
\usepackage{amssymb,amsmath,array}

\def\lsim{\raise0.3ex\hbox{$<$\kern-0.75em\raise-1.1ex\hbox{$\sim$}}}
\def\gsim{\raise0.3ex\hbox{$>$\kern-0.75em\raise-1.1ex\hbox{$\sim$}}}
\def\noi{\noindent}

\def\bei{\begin{itemize}}
\def\ei{\end{itemize}}
\def\bea{\begin{eqnarray}}
\def\eea{\end{eqnarray}}
\def\beqa{\begin{eqnarray}}
\def\eqa{\end{eqnarray}}
\def\beas{\begin{eqnarray*}}
\def\eeas{\end{eqnarray*}}
\def\beqas{\begin{eqnarray*}}
\def\eqas{\end{eqnarray*}}
\def\beq{\begin{equation}}
\def\eq{\end{equation}}
\def\eeq{\end{equation}}
\def\beqd{\begin{displaymath}}
\def\eeqd{\end{displaymath}}
\def\eqd{\end{displaymath}}

\def\beeq{\begin{eqnarray}} \def\eeeq{\end{eqnarray}}

\def\bef{\begin{frame}}

\def\slashchar#1{\setbox0=\hbox{$#1$}
   \dimen0=\wd0
   \setbox1=\hbox{/} \dimen1=\wd1
   \ifdim\dimen0>\dimen1
      \rlap{\hbox to \dimen0{\hfil/\hfil}}
      #1
   \else
      \rlap{\hbox to \dimen1{\hfil$#1$\hfil}}
      /
   \fi}

\newcommand{\scalingm}{2.2cm}

\newcommand{\scalingml}{2.6cm}
\newcommand{\scaling}{2.98cm}
\newcommand{\scalingb}{3.2cm}

\newcommand{\scalingL}{5cm}
\newcommand{\scalingA}{2.3cm}
\newcommand{\scalingB}{2.8cm}
\newcommand{\scalingC}{3.2cm}
\newcommand{\scalingD}{2.9cm}
\newcommand{\scalinglHTDA}{3cm}

\newcommand{\rb}{\underline{r}}
\newcommand{\kb}{\underline{k}}

\newcommand{\fin}{\end{document}}

\begin{document}
\title{QCD Factorizations in Exclusive $\gamma^* \gamma^* \to \rho^0_L \rho^0_L$}
%%%%   Paper title goes here  %%%%%%%%%%%%%%

%% 
%***********************************************************************
% AUTHORS INFORMATION AREA
%***********************************************************************
\author{ B.~Pire$^1$, M.~Segond$^{2,3}$, L.~Szymanowski$^{1,4}$
% Optional short acknowledgment: remove next line if non-needed
\thanks{Supported by the Polish Grant 1 P03B 028 28.} \,, S.~Wallon$^3$
% DO NOT MODIFY THE FOLLOWING '\vspace' ARGUMENT
\vspace{.3cm}\\
% Addresses and institutions (remove "1- " in case of a single institution)
1-  CPHT \\
\'Ecole Polytechnique, CNRS, Palaiseau, France 
%% Remove the next three lines in case of a single institution
\vspace{.1cm}\\
2- LPTHE \\
Universit\'e Paris 6 and 7, CNRS, Paris, France 
\vspace{.1cm}\\
3-  LPT\\ %\footnote{\,\,Unit{\'e} mixte 8627 du CNRS}, 
Universit\'e
Paris-Sud, CNRS, Orsay, France
\vspace{.1cm}\\
4- SINS \\
  Warsaw, Poland}

%%***********************************************************************
% END OF AUTHORS INFORMATION AREA
%***********************************************************************

\maketitle

\begin{abstract}
The exclusive process $e^+ \, e^- \to e^+ \, e^-  \, \rho^0_L \, \rho^0_L$ allows  to study 
various dynamics and factorization properties of perturbative QCD. At moderate energy,
we  demonstrate how
collinear QCD factorization emerges, involving either generalized
distribution amplitudes (GDA) or transition distribution amplitudes (TDA).
At higher energies,  in the Regge limit of QCD, we
 show that it offers a promising probe of the BFKL resummation effects to be studied at ILC.%the International Linear Collider (ILC). 
\end{abstract}

\section{Introduction: Exclusive processes at high energy QCD}
%\label{exclusive}
%\subsection{Motivation}

Since a decade, there has been much progress in experimental and theoretical  understanding of  hard exclusive processes, through the concepts of Generalized Parton Distribution and extensions.
%, including 
%Deeply Virtual Compton Scattering (involving Generalized Parton Distributions) and $\gamma \gamma$
%scattering
 %in fixed target 
  % $e^\pm p$ (HERMES, JLab, ...) experiments
%and at colliders, such as
 %  $e^\pm p$  (H1, ZEUS) or 
%$e^+ e^-$ (LEP, Belle, BaBar, BEPC).
Meanwhile, the  hard Pomeron \cite{bfkl} concept 
%and related resummed approaches  
has been developed and tested 
%at
% inclusive (total cross-section), semi-inclusive  (diffraction, forward jets, ...)
%and exclusive (meson production) level,
 for colliders at very large %center of mass 
energy.
%:
%   $e^\pm p$ (HERA), $p \bar{p}$ (Tevatron) and
%$e^+ e^-$ (LEP, ILC). 
The process 
%$p \bar{p}$ (TEVATRON: CDF, D0) and
%$e^+ e^-$ (LEP, ILC). We  focus on 
%a specific exclusive process: 
\beq
\label{processgg}
\gamma^* \gamma^* \to \rho^0_L \rho^0_L
\eq
with both $\gamma^*$ {\it hard}, in $e^+ e^- \!\to\! e^+ e^- \rho^0_L \rho^0_L$
with double tagged outoing leptons,
%It is a beautiful theoretical  laboratory for 
involves several {\it dynamical regions} (collinear, multiregge) and {\it factorization} properties of high energy QCD:
it allows a perturbative study of GPD-like objects at   moderate $s$ and of the hard Pomeron
at  asymptotic $s.$
%\subsection{From DIS to GDA and TDA in collinear factorization}
%Let us briefly recall the various type of extensions, from Deep Inelastic Scattering, which have been elaborated based on parton model, and their description in terms
%of factorization.
%Deep Inelastic Scattering, as an inclusive process, gives access 
%at the level of the cross section 
%to the forward amplitude through the optical theorem. Structure functions 
%can be written as 
%are 
%convolution of  (hard) Coefficient Functions with  (soft) Parton Distributions.  
Deeply Virtual Compton Scattering and meson electroproduction on a hadron  $\gamma^*h  \to \gamma \, h, \, h' \, h$,  as exclusive processes, give access to the full amplitude,\nopagebreak which is  a   convolution, for $-t \ll s,$  
of a (hard) CF
with a   (soft) Generalized Parton Distribution \cite{DM,GPD}.
Extensions were made from GPDs.  First \cite{DM,GDA}, the crossed process  $\gamma^* \, \gamma \to h \, h'$ can be factorized, for   $s \ll -t,$ as a  convolution of a (hard) CF  with a (soft) Generalized Distribution Amplitude  describing the correlator between two quark fields and a two hadron state.
\psfrag{r1}[cc][cc]{$\quad\rho(k_1)$}
\psfrag{r2}[cc][cc]{$\quad\rho(k_2)$}
\psfrag{p1}[cc][cc]{$\hspace{-1cm}\begin{array}{c} \!\! \ell_1 \hspace{.8cm}\slashchar{p}_1\\  \vspace{-.5cm}\hspace{.2cm}-\tilde{\ell}_1 \end{array}$}
\psfrag{p2}[cc][cc]{$\hspace{-1cm}\begin{array}{c} \!\! \vspace{.2cm}\ell_2 \\  \vspace{.55cm}\hspace{-.4cm}-\tilde{\ell}_2 \hspace{.75cm}\slashchar{p}_2 \end{array}$}
\psfrag{q1}[cc][cc]{$q_1$}
\psfrag{q2}[cc][cc]{$q_2$}
\psfrag{Da}[cc][cc]{DA}
\psfrag{HDA}[cc][cc]{$M_H$}
\psfrag{M}[cc][cc]{$M$}
\begin{figure}[h]
%\vspace{-.45cm}
\begin{center}
\hspace{-.4cm}
\scalebox{.63}
{
$\begin{array}{cccc}
\raisebox{-0.44 \totalheight}{\epsfig{file=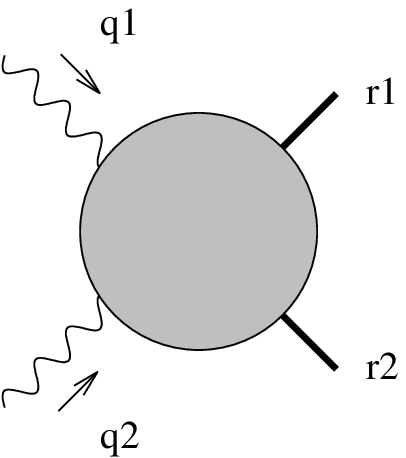,width=\scaling}}&=&
\raisebox{-0.44\totalheight}{\epsfig{file=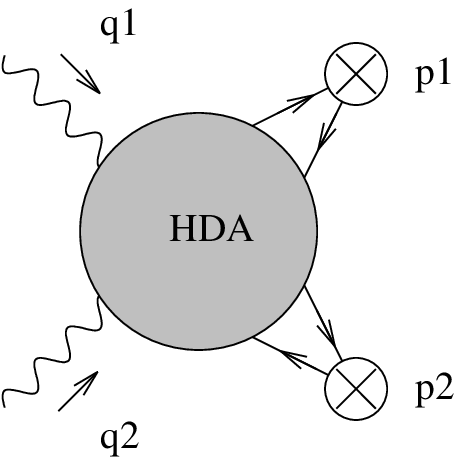,width=\scaling}}
& \begin{array}{c}
\raisebox{0.4 \totalheight}
{\psfrag{r}[cc][cc]{$\quad\rho(k_1)$}
\psfrag{pf}[cc][cc]{$\slashchar{p}_2$}
\epsfig{file=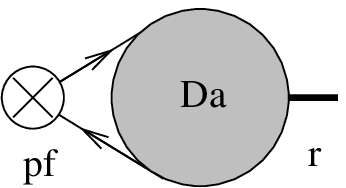,width=\scalingml}}\\
\raisebox{0.1 \totalheight}
{\psfrag{r}[cc][cc]{$\quad\rho(k_2)$}
\psfrag{pf}[cc][cc]{$\slashchar{p}_1$}
\epsfig{file=DA.eps,width=\scalingml}}
\end{array}
\end{array}
$}
\end{center}
\vspace{-.5cm}
\caption{\small %The amplitude of the process 
$\gamma^*(Q_1) \gamma^*(Q_2) \to \rho^0_L (k_{1})\rho^0_L(k_{2})$ with collinear factorization in $q \bar{q} \rho$ vertices.}%\vspace{-.7cm}
\label{Figfact}
\end{figure}
\noi
%hadron \, hadron$ 
Second \cite{TDA}, starting from  meson electroproduction and performing $t \leftrightarrow u$ crossing, and then allowing  the initial and the final hadron  to differ, we write the amplitude for the process $ \gamma^*  \, h \to h" \, h'$ 
%hadron' \, hadron".$ 
as  a  convolution of a  (hard) CF with a (soft) Transition  Distribution Amplitude describing the 
%$hadron \to hadron'$ 
$h \to h'$ transition and with a (soft) Distribution Amplitude (describing   $q \bar{q} h"$ vertex).
%of non perturbative nature, 
To describe the process  (\ref{processgg}), we rely on
 collinear factorization   at each  
 $q \bar{q} \rho$ vertex only. At high $Q_i^2$,  outgoing quarks are almost 
 collinear to the $\rho$ mesons, flying in the light cone direction 
$p_i$ (i=1 or 2) %and $p_2$, %(used as Sudakov vectors),
 and
 their momenta read
%\beqd
$\ell_i\sim z_i\, k_i$ and
$\tilde{\ell}_i\sim \bar{z}_i\, k_i  \,.$
%\eeqd
The amplitude $M$ is  factorized   as a convolution of a hard part $M_H$ with two $\rho^0_L$ DAs $\phi(z)$ (see Fig.\ref{Figfact}),  defined 
as matrix elements of non local quarks fields correlator   on the light 
cone (limiting ourselves to longitudinally polarized mesons to avoid potential end-point singularities). %\footnote{$\phi(z)=6 z (1-z)\,(1+ \sum_{n=1}^{\infty} a_{2  n} C^{3/2}_{2\,n} (2 z-1))\,.$ } 
%\vspace{-.2cm}
%$$%\vspace{-.7cm}\hspace{-2cm}
%\langle \rho^0_L(k)|\bar q(x)\gamma^\mu q(0)|0\rangle \!=\!
%\frac{f_\rho}{\sqrt{2}} k^\mu \!\!\!\int\limits_0^1 \! \! dz\,e^{iz(kx)}\phi(z)\;.% \;\;\;\;\;\mbox{for}\;\;q=u,d\;.
%\langle 0| \bar q(x)\gamma^\mu q(-x)|\rho_L(p)= \bar{q} q \rangle = f_{\rho}\,  p^\mu \!\!
%\int\limits_0^1 dz \, e^{i(2z-1)(px)}\phi(z)\,.
%$$
%We will now investigate $M_H$ in two regimes, the moderate and the high energy regime.

\section{Revealing QCD factorization at fixed $W^2$}

\subsection{Direct calculation}
%\begin{figure}[h]
%\begin{eqnarray*}
%\psfrag{q1}[cc][cc]{$q_1$}
%\psfrag{q2}[cc][cc]{$q_2$}
%\psfrag{l1}[cc][cc]{\lu}
%\psfrag{l1t}[cc][cc]{-\lut}
%\psfrag{l2}[cc][cc]{\ld}
%\psfrag{l2t}[cc][cc]{$\!-\tilde{\ell}_2$}
%\psfrag{n}[cc][cc]{}
%\vspace{-1.3cm}
%\centerline{\raisebox{1.3 \totalheight}{\scalebox{.43}{$
%\begin{array}{cccccccc}
%\raisebox{-0.44 \totalheight}{\epsfig{file=al1.eps,width=\scalingb}}
%&+&\raisebox{-0.44 \totalheight}{\epsfig{file=al2.eps,width=\scalingb}} &+&\raisebox{-0.44 \totalheight}{\epsfig{file=al3.eps,width=\scalingb}}&+&\raisebox{-0.44 \totalheight}{\epsfig{file=al4.eps,width=\scalingb}}
%\end{array}
%$}}}
%\vspace{-1.8cm}
%\end{eqnarray*}
%\caption{Diagrams contributing to $M_H^{\gamma^*_L \gamma^*_L}$
% \label{longDiagrams}}\vspace{-.7cm}
%\end{figure}

We compute \cite{gdatda} the amplitude $M$ following the Brodsky, Lepage approach \cite{BLphysrev24}, in the
 forward case for simplicity. It 
%At  Born order (quark exchange) and in the
% forward case\cite{gdatda}, 
%for simplicity,
%${\cal M}$ 
reads\footnote{$g^{\mu\,\nu}_T \equiv g^{\mu \nu} - \frac{p_1^\mu p_2^\nu + p_1^\nu p_2^\mu}{p_1 . p_2}\,;\quad s \equiv 2 \, p_1 \cdot p_2\,.$}, 
%following the seminal approach of\cite{BLphysrev24},
%is expressed as the sum of {\it two} tensors
%\beq
%\label{defM}
%${\cal M} = T^{\mu\, \nu}\epsilon_\mu(q_1)\epsilon_\nu(q_2)
%$
%\eq
%($g^{\mu\,\nu}_T=g^{\mu \nu} - (p_1^\mu p_2^\nu + p_1^\nu p_2^\mu)/p_1 . p_2$)
\beq
%\label{defT}
%{\cal M}
M=  T^{\mu\, \nu} \epsilon_\mu(q_1)\epsilon_\nu(q_2)\,,
\eq
which is expressed as the sum of {\it two} tensors
\beqa
\label{decT}
 && T^{\mu\, \nu}  = \frac{1}{2}g^{\mu\,\nu}_T\;  T^{\alpha\, \beta}g_{T\,\alpha\,\beta} \\
&&\hspace{-.7cm}+ \left(\!p_1^\mu +\frac{Q_1^2}{s}p_2^\mu\!\right)\!\left(\!p_2^\nu 
+ \frac{Q_2^2}{s}p_1^\nu\!\right)\!\frac{4}{s^2}  T^{\alpha\, \beta}p_{2\,\alpha}\, p_{1\,\beta}\,. \nonumber
\eqa
In the case of longitudinally  polarized photons and
at  Born order (quark exchange),
%their polarization vectors read
%\beq
%\label{polL}
%\epsilon_\parallel(q_1)=\frac{q_1}{Q_1} + \frac{2 Q_1}{s}p_2\,; \epsilon_\parallel(q_2)=\frac{q_2}{Q_2} + \frac{2 Q_2}{s}p_1\,.
%\eeq
%Due to QED gauge invariance, the first terms in RHS of (\ref{polL}) do  not contribute.
 this results in
 %the number of diagrams  reduces to 4, and equals %as illustrated in Fig.\ref{longDiagrams}. They result into
\beqa
&&T^{\alpha\, \beta}p_{2\,\alpha} \,p_{1\,\beta}= -\frac{s^2 f_\rho^2 C_F e^2 g^2(Q_u^2 +Q_d^2)}{8N_{c}Q_1^2 Q_2^2} \nonumber\\
&&\hspace{-.5cm}
\times 
\int\limits_0^1\,dz_1\,dz_2\,\phi(z_1)\,\phi(z_2) \left\{ \frac{1}{z_2 \bar z_1}
  \right. \\
&&\hspace{-.65cm} \left.+  \frac{(1-\frac{Q^2_1}{s})(1-\frac{Q^2_2}{s})}{(z_1+\bar z_1 \frac{Q^2_2}{s})(z_2+\bar z_2\frac{Q^2_1}{s})}\!+\! \left(\begin{array}{c} z_1 \leftrightarrow \bar z_1\\ z_2 \leftrightarrow \bar z_2\end{array}\right)\nonumber
 \right\}.
\label{resML}
\eqa
%\nopagebreak[4]
%\begin{eqnarray}
%&& T^{\alpha\, \beta}p_{2\,\alpha} \,p_{1\,\beta}= -\frac{s^2 f_\rho^2 C_F e^2 g^2(Q_u^2 +Q_d^2)}{8N_{c}Q_1^2 Q_2^2}
%\int\limits_0^1\,dz_1\,dz_2\,\phi(z_1)\,\phi(z_2) \nonumber
% \\
%&&\times \left\{\frac{(1-\frac{Q^2_1}{s})(1-\frac{Q^2_2}{s})}{(z_1+\bar z_1 \frac{Q^2_2}{s})(z_2+\bar z_2\frac{Q^2_1}{s})}
%+ \frac{(1-\frac{Q^2_1}{s})(1-\frac{Q^2_2}{s})}{(\bar z_1+ z_1 \frac{Q^2_2}{s})(\bar z_2+ z_2\frac{Q^2_1}{s})}
%+ \frac{1}{z_2 \bar z_1} + \frac{1}{z_1 \bar z_2} \right\}. \ \ \ \ \
%\label{resML}
%\end{eqnarray}
\noi
For transversally polarized photons, one gets\pagebreak
 %no simplification occurs and the 12 diagrams give
%\pagebreak
%, as illustrated in Fig.\ref{transDiagrams}.
%\vspace{-1cm}
 \begin{eqnarray}
\vspace{-1cm}
&&T^{\alpha\, \beta}g_{T\,\alpha \,\beta}= -\frac{e^2(Q_u^2 +Q_d^2)\,g^2\,C_F\,f_\rho^2}{4\,N_c\,s} \nonumber \\
&&\hspace{-.75cm}\times\!\!\int\limits_0^1\!\!dz_1\,dz_2\,\phi(z_1)\,\phi(z_2)\!\left\{\!2\left(\!1\!-\!\frac{Q^2_2}{s}\!\right)\!\!\left(\!1\!-\!\frac{Q^2_1}{s}\!\right) \right.\nonumber
 \\
&&\hspace{-.75cm}\times \! \left. \left[\!
\frac{1}{\!\left(z_2+\bar z_2\frac{Q_1^2}{s}\!\right)^2\!\left(\!z_1+\bar z_1\frac{Q_2^2}{s}\! \right)^2} \!+\!
\!\left(\!\begin{array}{c}\!\! z_1 \leftrightarrow \bar z_1\\ \!\!z_2 \leftrightarrow \bar z_2\end{array}\!\!\right)
 \right] \right. \nonumber \\
&&\hspace{-.75cm} \left. + \left(\frac{1}{\bar z_2\,z_1}- \frac{1}{\bar z_1\,z_2}  \right) 
\left[ \frac{1}{1-\frac{Q^2_2}{s}}\left( \frac{1}{\bar z_2+ z_2\frac{Q_1^2}{s}}\right.\right.\right. \nonumber \\
&&\hspace{-.75cm}\left. \left. \left.-  \frac{1}{ z_2+ \bar z_2\frac{Q_1^2}{s}}   \right)  -
\left(\begin{array}{c} z_1 \leftrightarrow  z_2\\ Q_1 \leftrightarrow Q_2 \end{array}\right)
 \right]
\right\} \,.
\label{resMT}
\end{eqnarray}
%As expected, since
%$Q_1^2$ and $Q_2^2$ are non-zero and  DAs vanishes at $z_i=0,$
%one encounter no end-point singularity in the $ z_i$ integration.
The $ z_i$ integrations have no end-point singularity ($Q_i^2 \neq 0$  and   $\phi(0)=0$).

%\subsection{Revealing QCD factorization}
\subsection{GDA for transverse photon in the limit $\Lambda_{QCD}^2 \ll W^2 \ll Max(Q_1^2,Q_2^2)$}

\begin{figure}[h]
\vspace{-.2cm}
%\begin{wrapfigure}{r}{1\columnwidth}
\centerline{\scalebox{1.}{\begin{tabular}{c}
%\centerline{
\scalebox{.52}
{$\begin{array}{cccc}
\raisebox{-0.44 \totalheight}
{\psfrag{r}[cc][cc]{$\quad\rho(k_1)$}
\psfrag{pf}[cc][cc]{$\slashchar{p}_2$}
\epsfig{file=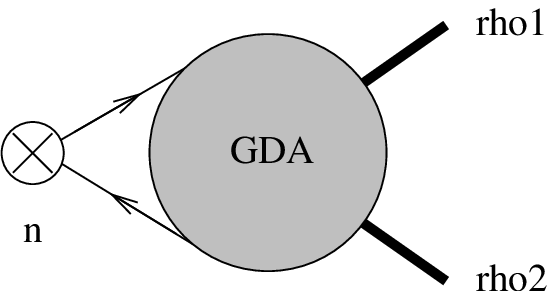,width=\scalingB}}&=&
\raisebox{-0.43\totalheight}{
\psfrag{GDA}[cc][cc]{$GDA_H$}
\epsfig{file=GDA.eps,width=\scalingC}}& \begin{array}{c}
\raisebox{0.51 \totalheight}
{\psfrag{r}[cc][cc]{$\quad\rho(k_1)$}
\psfrag{pf}[cc][cc]{$\slashchar{p}_2$}
\epsfig{file=DA.eps,width=\scalingA}}\\
\raisebox{-0.0 \totalheight}
{\psfrag{r}[cc][cc]{$\quad\rho(k_2)$}
\psfrag{pf}[cc][cc]{$\slashchar{p}_1$}
\epsfig{file=DA.eps,width=\scalingA}}
\end{array}
\end{array}$
}
%}
\\
\\
%\centerline{
\scalebox{.52}{
$\begin{array}{ccccccc}
\psfrag{GDA}[cc][cc]{$GDA_H$}
\psfrag{p1}[cc][cc]{$\slashchar{p}_1$}
\psfrag{p2}[cc][cc]{$\slashchar{p}_2$}
\psfrag{n}[cc][cc]{$\slashchar{n}$}
\raisebox{-0.43\totalheight}{\hspace{.1cm}\epsfig{file=GDA.eps,width=\scalingC}} &\hspace{-.2cm} =&\hspace{-.2cm}
\psfrag{p1}[cc][cc]{$\slashchar{p}_1$}
\psfrag{p2}[cc][cc]{$\slashchar{p}_2$}
\psfrag{n}[cc][cc]{$\slashchar{n}$}
\raisebox{-0.44 \totalheight}{\epsfig{file=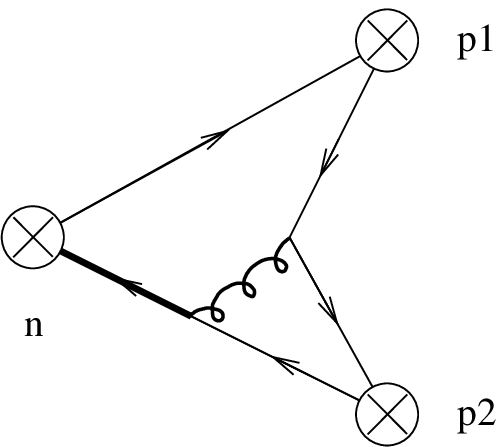,width=\scaling}}
&\hspace{-.2cm}+&\hspace{-.2cm}
\psfrag{p1}[cc][cc]{$\slashchar{p}_1$}
\psfrag{p2}[cc][cc]{$\slashchar{p}_2$}
\psfrag{n}[cc][cc]{$\slashchar{n}$}
\raisebox{-0.44\totalheight}{\epsfig{file=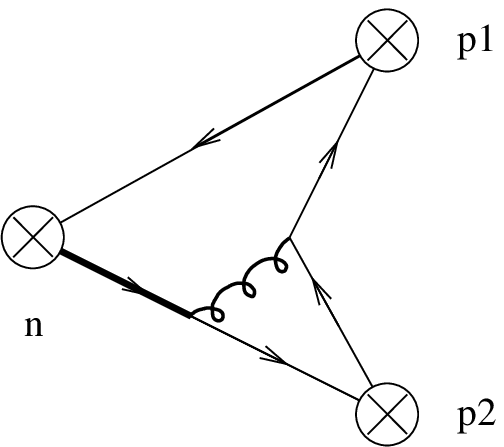,width=\scaling}}&\hspace{-.2cm}+&\hspace{-.2cm}
\psfrag{p1}[cc][cc]{$$}
\psfrag{p2}[cc][cc]{$$}
\raisebox{-0.44\totalheight}{\epsfig{file=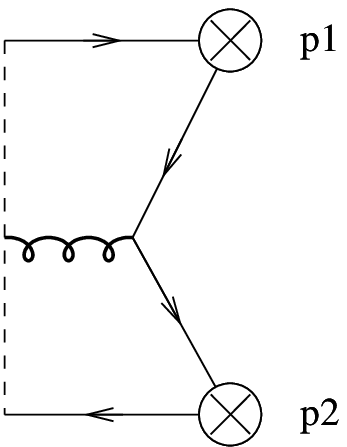,width=2cm}}
\end{array}$
}
%}
\end{tabular}}}\vspace{-.1cm}
\caption{Perturbative GDA factorization.}
\label{FigfactGDAH}
\end{figure}

\begin{figure}[h]
\vspace{-.8cm}
\centerline{\scalebox{.48}
{
$\begin{array}{ccccc}
\raisebox{-0.46 \totalheight}{\epsfig{file=THT.eps,width=\scaling}}
&=&
\raisebox{-0.46 \totalheight}{\epsfig{file=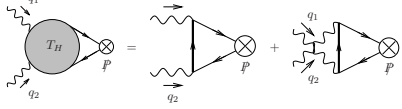,width=\scaling}}
&+&
\raisebox{-0.46\totalheight}{\epsfig{file=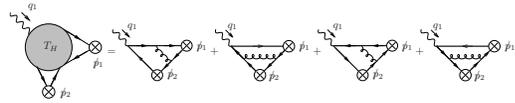,width=\scaling}}
\end{array}
$}}
\caption{Hard part $T_H$ at lowest order.} %(bold lines correspond to quark propagators)
\label{FigTHGDA}\vspace{-.5cm}
\end{figure}

\psfrag{r1}[cc][cc]{$\quad\rho(k_1)$}
\psfrag{r2}[cc][cc]{$\quad\rho(k_2)$}
\psfrag{p1}[cc][cc]{$\slashchar{p}_1$}
\psfrag{p2}[cc][cc]{$\slashchar{p}_2$}
\psfrag{p}[cc][cc]{$\qquad\slashchar{P}\qquad \slashchar{n}$}
\psfrag{n}[cc][cc]{}
\psfrag{q1}[cc][cc]{$q_1$}
\psfrag{q2}[cc][cc]{$q_2$}
\psfrag{GDA}[cc][cc]{${GDA}_H$}
\psfrag{Da}[cc][cc]{DA}
\psfrag{HDA}[cc][cc]{$M_H$}
\psfrag{M}[cc][cc]{$M$}
\psfrag{Th}[cc][cc]{$T_H$}
%\begin{equation*}
%\begin{figure}[htp]
\begin{figure}[h]
\begin{center}
\vspace{-.2cm}
\hspace{-.2cm}
\scalebox{.57}
{$\begin{array}{cccc}
\begin{array}{cccc}
\!\!\raisebox{-0.44\totalheight}{\epsfig{file=HDA.eps,width=\scalingD}}& 
\begin{array}{ccc}\raisebox{0.51 \totalheight}
{\psfrag{r}[cc][cc]{$\quad\rho(k_1)$}
\psfrag{pf}[cc][cc]{$\slashchar{p}_2$}
\epsfig{file=DA.eps,width=\scalingA}}\\
\raisebox{0. \totalheight}
{\psfrag{r}[cc][cc]{$\quad\rho(k_2)$}
\psfrag{pf}[cc][cc]{$\slashchar{p}_1$}
\epsfig{file=DA.eps,width=\scalingA}}\end{array}& &
\end{array}\\
\begin{array}{cccc}
=& \!
\raisebox{-0.44 \totalheight}
{\psfrag{r}[cc][cc]{$\quad\rho(k_1)$}
\psfrag{pf}[cc][cc]{$\slashchar{p}_2$}
\epsfig{file=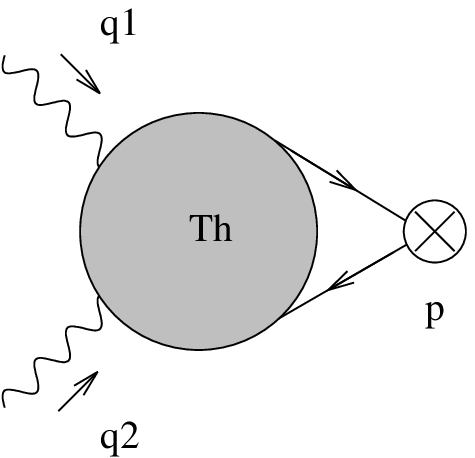,width=\scalingB}}&
\raisebox{-0.43\totalheight}{\epsfig{file=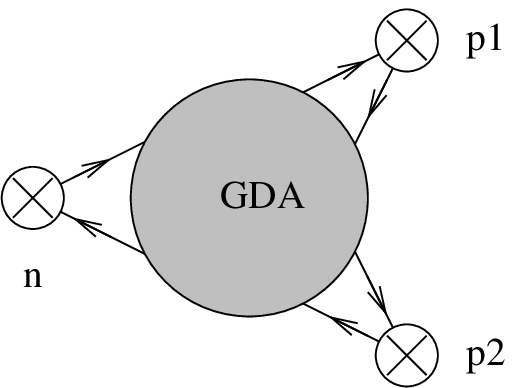,width=\scalingC}}& \begin{array}{c}
\raisebox{0.51 \totalheight}
{\psfrag{r}[cc][cc]{$\quad\rho(k_1)$}
\psfrag{pf}[cc][cc]{$\slashchar{p}_2$}
\epsfig{file=DA.eps,width=\scalingA}}\\
\raisebox{-0.0 \totalheight}
{\psfrag{r}[cc][cc]{$\quad\rho(k_2)$}
\psfrag{pf}[cc][cc]{$\slashchar{p}_1$}
\epsfig{file=DA.eps,width=\scalingA}}
\end{array}
\end{array}
\end{array}
$}
\end{center}
\vspace{-.4cm}
\caption{Factorisation of the  amplitude in terms of a GDA.}
\label{FigconvGDA}
\end{figure}
\noi When $W^2$ is smaller than the highest photon virtuality (for example $Q_1^2$),
 (\ref{resMT}) simplifies in\footnote{We denote $C=\frac{e^2(Q_u^2 +Q_d^2)\,g^2\,C_F\,f_\rho^2}{4\,N_c}$}
%in\footnote{In
%this example
%$\frac{W^2}{Q_1^2}=
%\frac{s}{Q_1^2}\left(1-\frac{Q_1^2}{s}
%\right)\left(1-\frac{Q_2^2}{s}  \right) \approx 1-\frac{Q_1^2}{s}\ll
%1;
\beqa
\label{simpMT}
&&\hspace{-.7cm}T^{\alpha\, \beta}g_{T\,\alpha \,\beta} \approx \frac{C}{W^2}
\int\limits_0^1\,dz_1\,dz_2\,\left(\frac{1}{\bar z_1+ z_1\frac{Q_2^2}{s}}
\right.\nonumber\\
&&\hspace{-1.05cm}\left. - \frac{1}{ z_1+ \bar z_1\frac{Q_2^2}{s}}   \right)\!\!
\left(\frac{1}{\bar z_2\,z_1} - \frac{1}{\bar z_1\,z_2}  \right)\! \phi(z_1)\,\phi(z_2),
% \frac{1}{1-\frac{Q^2_1}{s}}
\eqa
showing that the hard amplitude $M_H$ can be factorized as a convolution between a hard coefficient function $T_H$ and a ${GDA}_H,$ itself {\it perturbatively} computable (Fig.\ref{FigconvGDA}), extending the results of \cite{DFKV}.
This is proven at Born order. First one computes perturbatively
 the GDA  from its definition
as a bilocal correlator:
  $W^2$ being hard, the GDA can be factorized as  DA $\otimes$ $GDA_H$  $\otimes$ DA 
(Fig.\ref{FigfactGDAH}). A QCD Wilson line (last term in  Fig.\ref{FigfactGDAH}) has to be included to fulfil gauge invariance. It  vanishes in forward kinematics.
Second, one  computes the Born order hard part
%involves two contributions (see Fig.\ref{FigTHGDA}) and equals, for one flavored quark,
 (Fig.\ref{FigTHGDA}). These two results combine according to  Eq.(\ref{simpMT}).
\psfrag{r1}[cc][cc]{$\quad\rho(k_1)$}
\psfrag{r2}[cc][cc]{$\quad\rho(k_2)$}
\psfrag{p1}[cc][cc]{$\slashchar{p}_1$}
\psfrag{p2}[cc][cc]{$\slashchar{p}_2$}
\psfrag{p}[cc][cc]{$\qquad\slashchar{P}\qquad \slashchar{n}$}
\psfrag{n}[cc][cc]{}
\psfrag{q1}[cc][cc]{$q_1$}
\psfrag{q2}[cc][cc]{$q_2$}
\psfrag{GDA}[cc][cc]{$GDA$}
\psfrag{Da}[cc][cc]{DA}
\psfrag{HDA}[cc][cc]{$M_H$}
\psfrag{M}[cc][cc]{$M$}
\psfrag{Th}[cc][cc]{$T_H$}
\psfrag{rho1}[cc][cc]{$\quad\rho(k_1)$}
\psfrag{rho2}[cc][cc]{$\quad\rho(k_2)$}
%\begin{equation*}
%\end{wrapfigure}
%%%%%
\psfrag{p}[cc][cc]{$\slashchar{P}$}
\psfrag{q1}[cc][cc]{$q_1$}
\psfrag{q2}[cc][cc]{$q_2$}
\psfrag{Th}[cc][cc]{$T_H$}
\psfrag{n}[cc][cc]{$\slashchar{n}$}
%\[
%T_H(z) = -4\,e^2\,N_c\,Q_q^2\,\left( \frac{1}{\bar z +z\,\frac{Q_2^2}{s}} -
%   \frac{1}{z + \bar z\,\frac{Q_2^2}{s}} \right)\,.
%\]
%\begin{equation*}

%This ends the proof 
 %that $T^{\alpha\, \beta}g_{T\,\alpha \,\beta}$ factorizes into $Hard \, part \, \otimes \, GDA,$ 
%\beq
%\label{factGDAfinal}
%T^{\alpha\, \beta}g_{T\,\alpha \,\beta}=
%\frac{ e^2}{2}\left(Q_u^2 +Q_d^2  \right)\int\limits_0^1 \,dz\,\left(\frac{1}{\bar z +z \frac{Q_2^2}{s}}
%- \frac{1}{ z +\bar z \frac{Q_2^2}{s}}  \right)\Phi^{\rho_L\rho_L}(z,\zeta\approx 1,W^2)
%\eq

%This is a limiting case of the original equation obtained by {\aut D. Müller et al} (2000)
%It extends the studies of $\gamma^* \gamma^* \to \pi \pi$ by {\aut M. Diehl et al} (2000) 
%We limited ourselves to the case of $t=t_{min}$ 

\subsection{TDA for longitudinal photon in the limit $Q_1^2 \gg Q_2^2$ (or $Q_1^2 \ll Q_2^2$)}

\begin{figure}[h]
\psfrag{q1}[cc][cc]{$q_1$}
\psfrag{q2}[cc][cc]{$q_2$}
\psfrag{p1}[cc][cc]{$\slashchar{p}_1$}
\psfrag{p2}[cc][cc]{$\slashchar{p}_2$}
\psfrag{n}[cc][cc]{$\slashchar{p}_1$}
\psfrag{p}[cc][cc]{$\slashchar{p}_2$}
\psfrag{Tda}[cc][cc]{$TDA_H$}
\psfrag{Th}[cc][cc]{$T_H$}
\psfrag{Da}[cc][cc]{DA}
%\begin{equation*}
\vspace{-.1cm}
%\hspace{1cm}
\hspace{1cm}\scalebox{.6}
{$\begin{array}{c}
\begin{array}{cc}
\raisebox{-0.44 \totalheight}{\epsfig{file=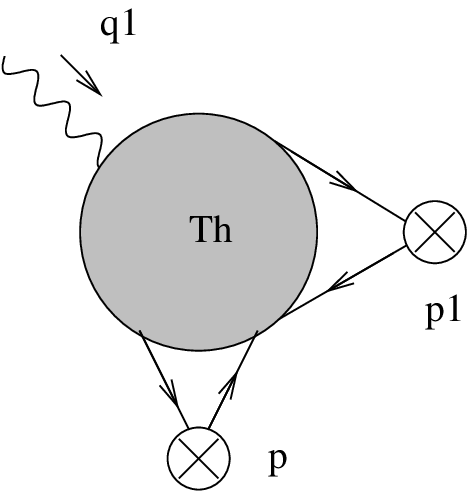,width=\scaling}}&
\raisebox{-0.25\totalheight}
{\psfrag{r}[cc][cc]{$\qquad \rho(k_1)$}
\psfrag{pf}[cc][cc]{\raisebox{-1.05 \totalheight}{$\slashchar{p}_2$}}\epsfig{file=DA.eps,width=\scalingm}}
\end{array}\\
\\
\begin{array}{cc}
\raisebox{-0.44 \totalheight}{\epsfig{file=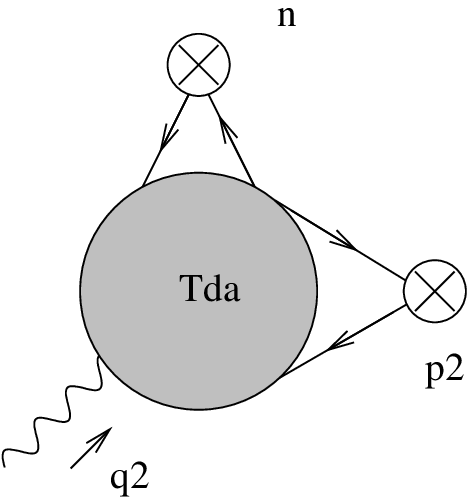,width=\scaling}}&
\raisebox{-0.54\totalheight}
{\psfrag{r}[cc][cc]{$\qquad \rho(k_2)$}
\psfrag{pf}[cc][cc]{\raisebox{-1.05 \totalheight}{$\slashchar{p}_1$}}\epsfig{file=DA.eps,width=\scalingm}}
\end{array}
\end{array}
$}
\caption{Factorization of the amplitude in terms of a TDA.}
\label{FigTDAfact} 
\end{figure}

\begin{figure}[h]
\psfrag{p}[cc][cc]{$\,\, \slashchar{p}_2$}
\psfrag{p1}[cc][cc]{$\,\, \slashchar{p}_1$}
\psfrag{p2}[cc][cc]{$\,\, \slashchar{p}_2$}
\psfrag{q1}[cc][cc]{$q_1$}
\psfrag{Th}[cc][cc]{$T_H$}
\psfrag{n}[cc][cc]{$\slashchar{n}$}
%\begin{equation*}
\vspace{-.2cm}
\hspace{-.15cm}
\centerline{\scalebox{.41}
{
$
\begin{array}{ccccccccc}
\raisebox{-0.46 \totalheight}{\epsfig{file=THL.eps,width=\scalinglHTDA}}
&\!\!\! =&\!\!\!\!\!\!\!\!
\raisebox{-0.46 \totalheight}{\epsfig{file=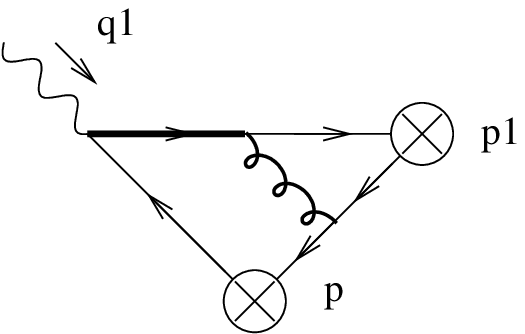,width=\scalinglHTDA}}
&\!\!\!+&\!\!\!\!\!\!\!\!
\raisebox{-0.46 \totalheight}{\epsfig{file=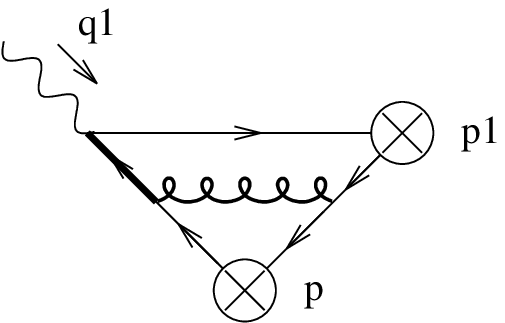,width=\scalinglHTDA}}
&\!\!\!+&\!\!\!\!\!\!\!\!
\raisebox{-0.46\totalheight}{\epsfig{file=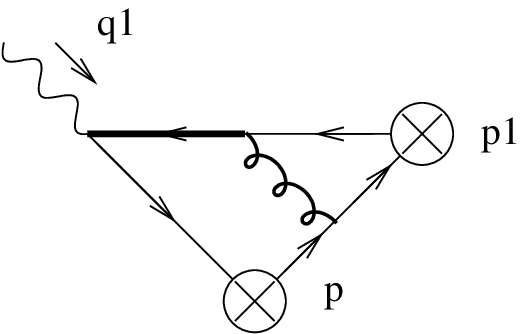,width=\scalinglHTDA}}
&\!\!\!+&\!\!\!\!\!\!\!\!
\raisebox{-0.46\totalheight}{\epsfig{file=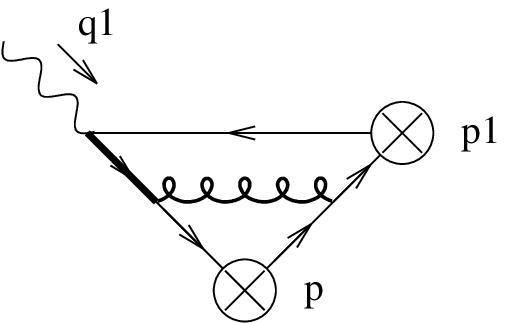,width=\scalinglHTDA}}
\end{array}
$}}
%\vspace{-.6cm}
\caption{Hard part $T_H$ at lowest order.}%\vspace{-1.6cm}
\label{HardTDA}
\end{figure}

The amplitude 
%${\cal M}
 (\ref{resML}) simplifies in this limit as  
\begin{eqnarray}
\label{simpML}
&&\hspace{-.6cm}T^{\alpha\,\beta}p_{2\,\alpha}p_{1\,\beta} =
-i \frac{C}{2}\int\limits_{-1}^1 dx\,\int\limits_0^1 dz_1\,
\left[\frac{1}{\bar z_1(x-\xi)} \right. \nonumber \\
&&\hspace{-.8cm}\left.+ \frac{1}{z_1(x+\xi)} \right]\! \phi(z_1)  N_c\!\! \left[\Theta(1\ge x \ge \xi)\, \phi\!\left(\frac{x-\xi}{1-\xi} \!\right) \right. \nonumber \\
&&\hspace{-.6cm} \left.-
\Theta(-\xi \ge x \ge -1) \, \phi\left( \frac{1+x}{1-\xi} \right) \right]\!,
\end{eqnarray}
to be interpreted 
%, in this limiting case $Q_1^2 \gg Q_2^2$ (or $Q_1^2 \ll Q_2^2$), 
as
a convolution $M =   TDA \otimes CF \otimes DA.$
The TDA is defined in the usual GPD kinematics, with 
% (see Fig.\ref{FigTDAkin}),
 skewedness  $\xi= Q_1^2/(2s - Q_1^2)$ and
 momentum fractions along $n_2 =\frac{p_2}{1+\xi}.$
This factorisation (Fig.\ref{FigTDAfact})  is proven at Born order. First, one computes perturbatively the TDA $\gamma^*_L \to \rho_L^0$ defined\pagebreak
% =\frac{p_2}{1+\xi}.$ 
%\beas
%&&\int\,\frac{dz^-}{2\pi s}\,e^{ix(n_2.z)}\,\langle \rho^q_L(k_2)|
%\bar q(-z/2)\,\slashchar{n_1}\,\exp\{-ieQ_q\!\int\limits_{z/2}^{-z/2}\,
%dy_\mu\,A^\mu(y) \}q(z/2)|\gamma^*(q_2)\rangle
%\nonumber \\
%&&= \frac{e\,Q_q\,f_\rho}{n_2^+}\,\frac{1}{Q_2^2}\,
%\epsilon_\nu(q_2)\left((1+\xi)n_2^\nu +\frac{Q_2^2}{s(1+\xi)}n_1^\nu\right)\,T(x,\xi,t_{min})\;,
%\eeas
%\vspace{-.2cm}
%\begin{figure}
\noi  by a bilocal correlator. 
\begin{figure}[h]
%}
%%%%%%%%%
%%%%%%%%%%%
%\centerline{
\vspace{-.2cm}
\begin{tabular}{c}
\psfrag{r1}[cc][cc]{$\quad\rho(k_1)$}
\psfrag{r2}[cc][cc]{$\quad\rho(k_2)$}
\psfrag{p1}[cc][cc]{$\slashchar{p}_1$}
\psfrag{p2}[cc][cc]{$\slashchar{p}_2$}
\psfrag{p}[cc][cc]{$\qquad\slashchar{P}\qquad \slashchar{n}$}
\psfrag{n}[cc][cc]{$\slashchar{p}_1$}
\psfrag{q1}[cc][cc]{$q_1$}
\psfrag{q2}[cc][cc]{$q_2$}
\psfrag{Tda}[cc][cc]{$TDA$}
\psfrag{Da}[cc][cc]{DA}
\psfrag{HDA}[cc][cc]{$M_H$}
\psfrag{M}[cc][cc]{$M$}
\psfrag{Th}[cc][cc]{$T_H$}
\psfrag{rho1}[cc][cc]{$\quad\rho(k_1)$}
\psfrag{rho2}[cc][cc]{$\quad\rho(k_2)$}
%\begin{equation*}
%\centerline{
\scalebox{.53}
{$\begin{array}{cccc}
\raisebox{-0.44 \totalheight}
{\psfrag{r}[cc][cc]{$\quad\rho(k_1)$}
\psfrag{pf}[cc][cc]{$\slashchar{p}_2$}
\epsfig{file=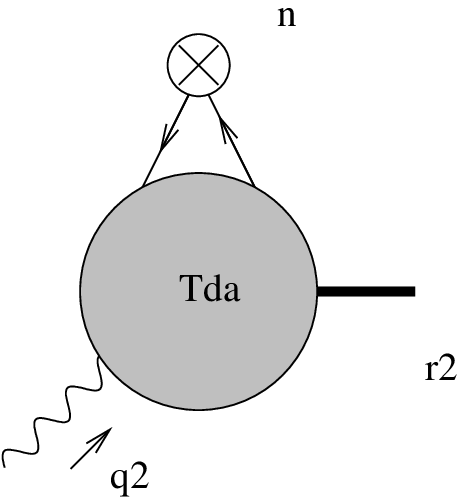,width=\scalingB}}&=&
\raisebox{-0.43\totalheight}{
\psfrag{Tda}[cc][cc]{$TDA_H$}
\epsfig{file=TDA.eps,width=\scalingC}}& \begin{array}{c}
\raisebox{0.51 \totalheight}
{\psfrag{r}[cc][cc]{$\quad\rho(k_1)$}
\psfrag{pf}[cc][cc]{$\slashchar{p}_2$}
}\\
\raisebox{-0.0 \totalheight}
{\psfrag{r}[cc][cc]{$\quad\rho(k_2)$}
\psfrag{pf}[cc][cc]{$\slashchar{p}_1$}
\epsfig{file=DA.eps,width=\scalingA}}
\end{array}
\end{array}$
}
\\
\hspace{-.3cm}
\vspace{.3cm}
\raisebox{-0.46 \totalheight}{\rm with}
\\
\vspace{.3cm}\hspace{-.3cm}
\psfrag{p}[cc][cc]{$\slashchar{p}_2$}
\psfrag{p2}[cc][cc]{$\slashchar{p}_2$}
\psfrag{q1}[cc][cc]{$q_1$}
\psfrag{q2}[cc][cc]{$q_2$}
\psfrag{Tda}[cc][cc]{$TDA_H$}
\psfrag{n}[cc][cc]{$\slashchar{p}_1$}
%\begin{equation*}
\scalebox{.5}
{
$\begin{array}{ccccccc}
\raisebox{-0.46 \totalheight}{\epsfig{file=TDA.eps,width=\scalingb}}
&\hspace{-.2cm}=&\hspace{-.5cm}
\raisebox{-0.46 \totalheight}{\epsfig{file=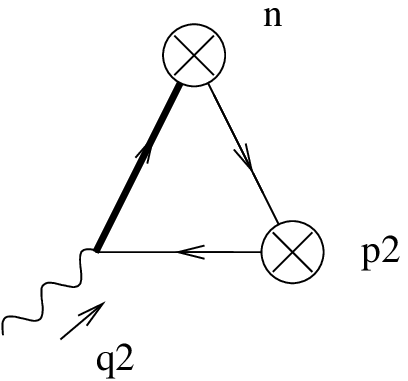,width=\scalingb}}
&\hspace{-.3cm}+&\hspace{-.5cm}
\raisebox{-0.46\totalheight}{\epsfig{file=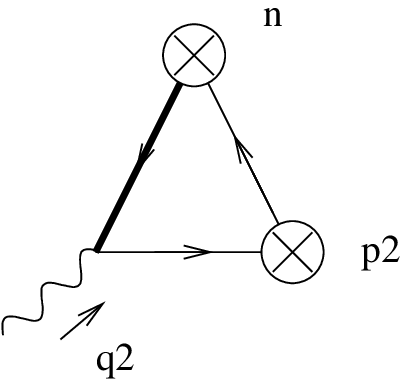,width=\scalingb}}
&\hspace{-.3cm}+&\hspace{-.5cm}
\raisebox{-0.42\totalheight}{\epsfig{file=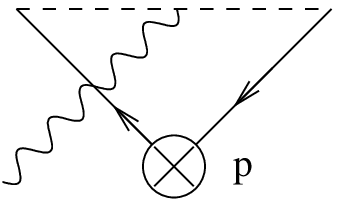,width=\scalingml}}
\end{array}
$}
\end{tabular}
\vspace{-.6cm}
\caption{Perturbative TDA factorization.}
\label{FigTDAfactPert}
\end{figure}
%\end{figure}
$Q_2^2$ being hard, the TDA  factorizes (Fig.\ref{FigTDAfactPert}).  
To satisfy gauge invariance, a QED Wilson line is included (last term of Fig.\ref{FigTDAfactPert}).  
Second, the Hard term is computed at Born order (see Fig.\ref{HardTDA}).
These two results combine according to  (\ref{simpML}).
\vspace{-.0cm}
%\vspace{-.4cm}
%\vspace{.2cm}

\section{The high energy limit}
%\subsection{Aims}

\noi QCD
in the perturbative  Regge limit  is governed by gluons. 
%(dominance of spin 1 exchange in
%$t$ channel).Visiting Fellow of FNRS (Belgium)
BFKL %(and extensions: NLL, saturations effects, ...) 
enhancement is expected to be important at large  rapidity. %(see Fig.\ref{FigBFKL}). 
The exclusive process (\ref{processgg}) tests this limit \cite{PSW,EPSW,SSW},
 for both $Q_i^2$   {\it hard} %(IR-safeness of the $\gamma^*$ probes) 
and of the {\it same order}. For $s_{\gamma^*\gamma^*}
  \gg -t, Q_1^2,Q_2^2,$ we rely on the impact representation
which reads\pagebreak%\vspace{-.1cm}
\beqas
&&\hspace{-.72cm}{\cal M} \!\!=\!\! \frac{is}{16\pi^4}\!\!\!\int\!\!\!\frac{d^2\,\kb}{\kb^2(\rb \!-\!\kb)^2}
\!{\cal J}^{\gamma^*_{L,T}(q_1) \to \rho^0_L(k_1)}\!(\kb,\rb\! -\!\kb)\nonumber\\
&&\hspace{-.7cm}\times {\cal J}^{\gamma^*_{L,T}(q_2) \to \rho^0_L(k_2)}(-\kb,-\rb +\kb) 
\eqas
at Born order
The impact factors ${\cal J}^{\gamma^*_{L,T}}$ are rational functions of  the transverse momenta $(\kb,\rb)$.
The  2-d 
integration is treated analytically, through
 conformal transformations. 
% (to suppress collinear dynamics à la DGLAVisiting Fellow of FNRS (Belgium)P\cite{dglap} and ERBL\cite{ERBL}).
\begin{figure}[h]
%\centerline{
%\hspace{-7cm}
\scalebox{.95}{
\begin{picture}(300,198)
\put(0,100){\epsfxsize=0.8\columnwidth{\centerline{\epsfbox{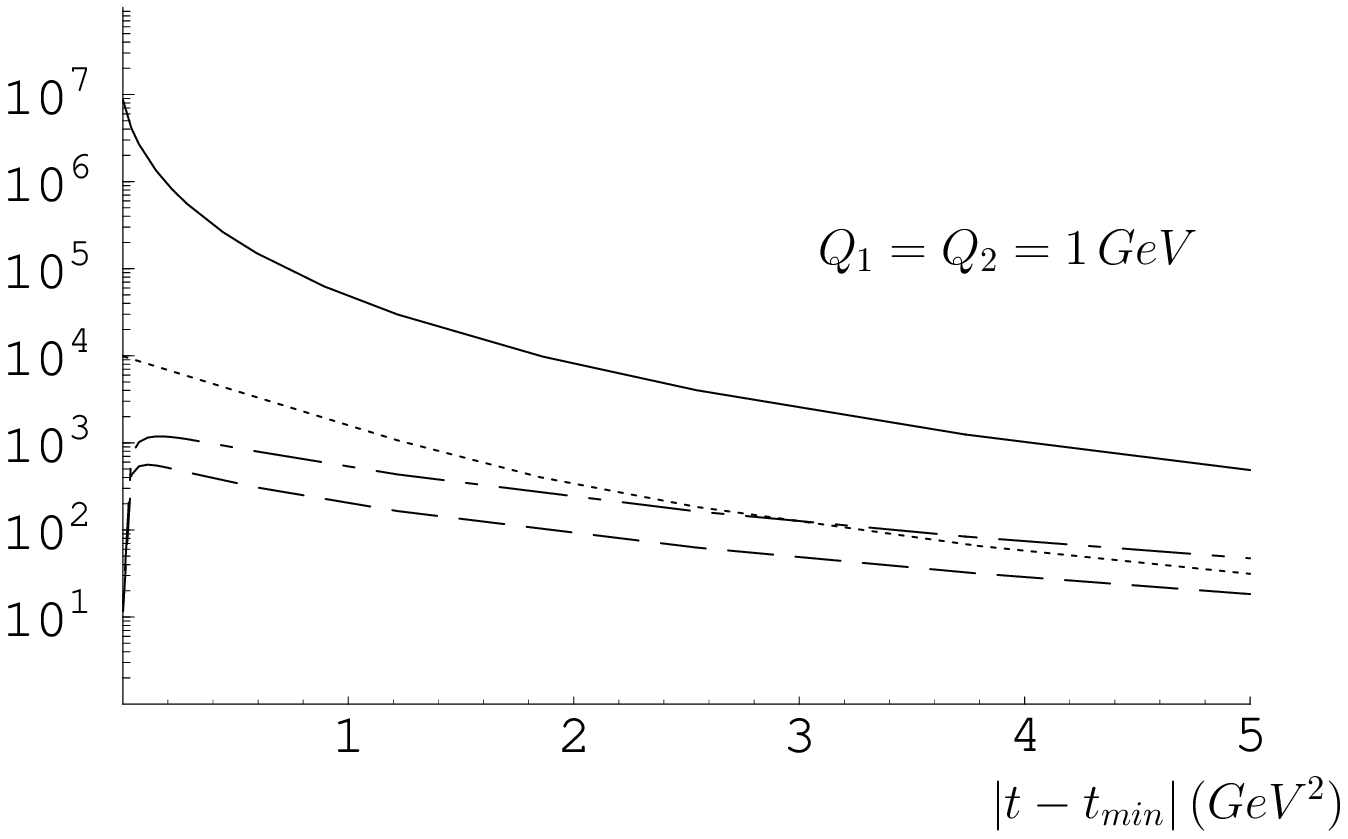}}}}
%\put(106,0){\epsfxsize=\wids{\centerline{\epsfbox{preparationFigsigmalarget.ps}}}}
\put(86,155){\tiny$LL$}
\put(86,143){\tiny$LT$}
\put(86,136){\tiny$T\neq T'$}
\put(86,127){\tiny$T= T'$}
\put(-2,2){\epsfxsize=0.85\columnwidth{\centerline{\epsfbox{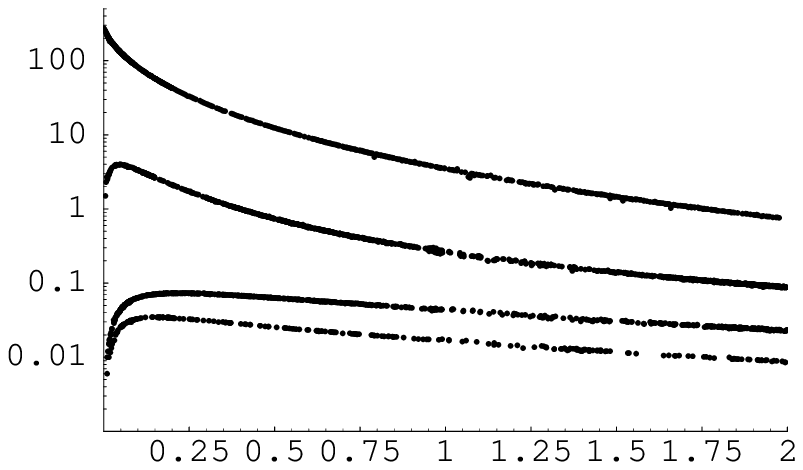}}}}
\put(86,74){\tiny $LL$}
\put(86,58){\tiny$LT$}
\put(86,44){\tiny$T\neq T'$}
\put(86,27){\tiny$T= T'$}
%\put(0,31000){ \large $\left.\frac{d \sigma^{\gamma^*_L\gamma^*_L\to \rho^0_L
      %\rho^0_L}}{d t}\right|_{t=t_{min}} \, (fb/GeV^2)$}
%\put(41500,1000){\large $Q_2^2/Q_1^2$}
%\put(40000,20000){ $Q_1 \, (GeV)$}
\put(128,1){\scalebox{.55}{ $|t-t_{min}|  \, (GeV^2)$}}
\put(22,102){\scalebox{.55}{$\frac{d\sigma^{e^+ e^- \to \, e^+ e^- \rho_L \rho_L}}{dt} (fb/GeV^2)$}}
\end{picture}
}\vspace{-.35cm}
%}
\caption{$\gamma^*_{L,T}\gamma^*_{L,T} \to  \rho_L^0  \;\rho_L^0$  (up) and $e^+e^- \to e^+e^- \rho_L^0  \;\rho_L^0$ (down) differential cross-sections.}\vspace{-.3cm}
\label{FigDiff}
\end{figure}
%\footnote{$\kb$ = Eucl. $\leftrightarrow $ $k_\perp$ = Mink.}
%(method inspired by  Vassiliev in 2-d coordinate space). 
%The idea is to reduce the number of propagators, in order to be able to perform standard Feynman parameter integration.
%the whole computation involves integrals with up to \alert{4 propagators (2 massive, with different masses)} which we would have been enable to compute without this method
The integrations over %momentum fractions 
$z_1$ and $z_2$ (hidden in ${\cal J}$) are performed numerically.
%We use $Q_1 Q_2$ as a scale for $\alpha_S$.
% (3 loops). 
Cross-sections are strongly peaked  at small $Q_i^2$ and  $t,$
and longitudinally polarized photons dominates (Fig.\ref{FigDiff}up).
% at small $t.$Visiting Fellow of FNRS (Belgium)
%\footnote{Any amplitude with at least one longitudinaly polarized phton vanishes at $t=t_{min}$.}. 
The non-forward Born order cross-section for
 $e^+e^- \to e^+e^- \rho_L^0  \rho_L^0$ is obtained using the 
 equivalent photon approximation. Defining 
 $y_{i}$ as the longitudinal momentum fractions of the bremsstrahlung photons, 
one finds that $\sigma^{e^+ e^- \to e^+ e^- \rho_L \rho_L}$  gets its main contribution from  the low $y$ and $Q^2$ region,  which is the very forward region.
At 
ILC, 
${\sqrt s_{e^+e^-}}  \! \!=\!\! 500\, {\rm GeV},$ 
 with $125 \,{\rm fb^{-1}}$ per year. The measurement seems feasible since
% within 4 years of running at $500 \,{\rm GeV},$
%including a possible scan in energy between $200 \,{\rm GeV}$ and $500 \,{\rm GeV}$
%and an upgrade at  $1\,{\rm TeV},$ with a luminosity of $1 \,{\rm ab^{-1}}$ within 3 to 4 years.
each detector design includes a very forward electromagnetic calorimeter %for luminosity measurement,
with tagging angle for outgoing leptons down to 5 mrad.
Fig.\ref{FigDiff}down displays our results within
the Large Detector Concept.
We obtain
%, at $\sqrt{s_{e^+e^-}}= 500$ GeV, 
$\sigma^{tot}= 34.1 \,{\rm fb}$ 
%\begin{eqnarray*}
%\sigma^{LL}&=& 32.4 \, {\rm fb} \\
%\sigma^{LT}&=& 1.5 \, {\rm fb} \\
%\sigma^{TT}&=& 0.2 \,{\rm fb}\\
%\sigma^{tot}&=& 34.1 \,{\rm fb} 
%\end{eqnarray*}
%which leads to
and $4.3 \,10^3$ events per year.
% with foreseen luminosity.
The  LL BFKL enhancement is enormous but not trustable, since it is well known that NLL  BFKL
is far below LL. Work  to implement  {\it resummed} LL  BFKL effects \cite{Khoze} 
%at $e^+e^-$ level 
is in progress, with results in accordance with the NLL based one \cite{IvanovPapa}. The obtained enhancement  is less dramatic ($\sim 5$) than with LL BFKL, but {\it still visible}.
%------------------------------------------------------------------------------
%       Bibliography
%------------------------------------------------------------------------------
%\fin

\begin{footnotesize}
\bibliographystyle{blois07}

\end{footnotesize}

\end{document}